\documentstyle[11pt,moriond,epsfig]{article}

\def\NPB{{\em Nucl. Phys.} {\bf B}}
\def\PLB{{\em Phys. Lett.}  {\bf B}}

\def\ZPC{{\em Z. Phys.} {\bf C}}

\def\be{\begin{equation}}
\def\ee{\end{equation}}
\def\bea{\begin{eqnarray}}
\def\eea{\end{eqnarray}}

\def\cA{{\cal{A}}}

\def\cK{{\cal{K}}}

\def\beq{\begin{equation}}   \def\eeq{\end{equation}}
\def\as{\alpha_s}
\def\eps{\epsilon}
\def\ee{e^+e^-}
\begin{document}
\vspace*{4cm}
\title{RENORMALON CONTRIBUTIONS TO FRAGMENTATION FUNCTIONS IN
NON SINGLET DIS}

\author{ M.DASGUPTA}

\address{Cavendish Laboratory, Madingley Road, Cambridge CB3 OHE, U.K}

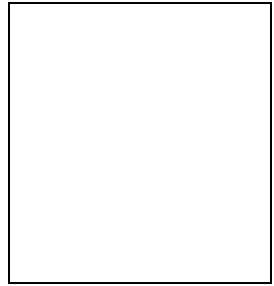
\maketitle\abstracts{This talk discusses the power behaved corrections
to fragmentation functions for the current jet in non-singlet deep
inelastic scattering. These corrections are estimated by means of a
renormalon model using a
dispersive approach \cite{BPY}. The assumption of a universal
infrared finite coupling enables us to provide quantitative estimates.}

\section{Introduction}
The experimental study of final state, single hadron, momentum
distributions is now being carried out at HERA \cite{fragexp}. Like the energy
distribution in $e^{+}e^{-}$, this quantity naturally depends on the parton to hadron
fragmentation functions. While it is not possible to calculate
fragmentation functions within perturbation theory, one can study their
evolution with $Q^2$ (the logarithmic scaling violation)
perturbatively and this
enables us to extract $\alpha_s$ from the data. However, reliance on a
purely perturbative approach still causes problems as it does not
account for higher twist pieces. Though these are suppressed by powers
of a large scale, their functional dependence on the phase space
parameters generally leads to sizable effects, even comparable in some
cases to
the size of the NLO term of the corresponding perturbative estimate. This would naturally mean that one must account
for these pieces before any comparison with data is made. 

The most reliable way in which one can model these effects is the
renormalon approach. Several papers have been written on these
techniques and their applications to structure functions \cite{DasWebDIS,Schafer}
fragmentation functions in $e^{+}e^{-}$ \cite{self}, and event shape
variables \cite{gavin,yuri}. The phenomenological success has been
encouraging enough to continue applying these techniques. 

For details of the renormalon technique we can refer to the review
talks in these proceedings \cite{volodya}. For our purposes we provide
a very brief introduction to the dispersive approach in section 2. In
section 3 we define the observables considered and explain our
notation. The calculations are described in section 4 and finally
quantitative results are presented in the form of a figure.    

\section{The Dispersive Approach}
Our estimate of the leading power corrections to the perturbative
results are based on the approach of reference \cite{BPY}. Non-perturbative effects at long distances are
assumed to give rise to a modification $\delta\alpha_{eff}(\mu^{2})$
in the QCD effective coupling at low values of the scale
$\mu^{2}$. The effect on some observable $F$ is then given by a
characteristic function ${\mathcal{F}} (x,\epsilon)$ as follows:
\begin{equation}
\delta F(x,Q^2) = \int_{0}^{\infty} \frac{d\mu^2}{\mu^{2}}
\delta\alpha_{eff}(\mu^2) \dot{\mathcal{F}}(x,\epsilon=\mu^2/Q^2)
\end{equation}
where 
\begin{equation}
\dot{\mathcal{F}}(x,\epsilon) = -\epsilon \frac{\partial}{\partial
\epsilon}{\mathcal{F}}(x,\epsilon)
\end{equation}
The characteristic function is obtained by computing the relevant
one-loop graphs with a non-zero gluon mass $\mu$.
\\
Arbitrary finite modifications of the effective coupling at low scales
would generally introduce power corrections of the form $1/k^{2p}$ into
the ultraviolet behaviour of the running coupling itself.
Such a modification would
destroy the basis of the operator product expansion.
This leads to the constraint that only terms in the small-$\epsilon$
behaviour of the characteristic function that are non-analytic at
$\epsilon=0$ will lead to power behaved non-perturbative contributions.
In fact we can choose to express our results only in terms of
$\alpha_s$ without the need to explicitly introduce $\alpha_{eff}$\cite{myself}.
Then the results can be expressed in terms of ordinary rather than the
logarithmic moment integrals of the coupling. The relevant moment
integral in our case (or more generally for any $\frac{1}{Q^2}$
correction) is 
\beq
\cA_2 \equiv \frac{C_F}{2\pi}
\int_0^\infty \frac{d\mu^2}{\mu^2}\,\mu^2\,\delta\as(\mu^2)\;.
\eeq
\section{Fragmentation in DIS}
For the usual reasons (absence of complications due to the remnant
jet, similarity with $e^{+}e^{-}$) one
chooses to work only in the current hemisphere of the Breit frame. For more details regarding
the Breit frame the reader is referred to \cite {myself} and references
therein. Since we wish to include only
particles in the current hemisphere, we define the fragmentation
function $F^h$ for  a given hadron
species as a function of the variable $z = 2p_h.q/Q^2$, which measures the fraction of the hadron's
momentum along the current direction and takes values $0 < z < 1$ in the current hemisphere.
The observable we wish to study is then given by
\beq
\label{bakwas}
F^{h} \left (z,x,Q^2 \right)=\frac{d^3\sigma^h}{dx dQ^2 dz}\bigg/ \frac{d^2\sigma^h}{dx dQ^2} 
\eeq
where $x=Q^2/2(P.q)$ if $P$ is the incoming hadron momentum. The denominator
of this expression is the fully inclusive deep-inelastic cross section
\beq\label{diffcsh}
\frac{d^2\sigma}{dx dQ^2} = \frac{2\pi\alpha^2}{Q^4}
\left\{\left[1+(1-y)^2\right] F_T(x)+2(1-y)
F_L(x)\right\}
\eeq
where $F_T(x)=2F_1(x)$ and $F_L(x) = F_2(x)/x
- 2F_1(x)$ are the usual transverse and longitudinal
structure functions. The numerator is given by
\beq
\frac{d^3\sigma^h}{dx dQ^2 dz} = \frac{2\pi\alpha^2}{Q^4}
\left\{\left[1+(1-y)^2\right] F^h_T(x,z)+2(1-y) F^h_L(x,z)\right\}
\eeq
where  $F^h_T(x,z)=2F^h_1(x,z)$ and $F_L^h(x,z) = F_2^h(x,z)/x
- 2F^h_1(x,z)$ are generalized transverse and longitudinal
structure functions.  The parton model yields $F^h_L(x,z)=F_L(x)=0$ and
\beq\label{Borncs}
F^h_T(x,z) = \sum_q e_q^2 [q(x)D^h_q(z)+\bar q(x)D^h_{\bar q}(z)]\;,
\eeq
while
\beq\label{Unborncs}
F^h_T(x)=\sum_q e_q^2 [q(x)+\bar q(x)] = f(x)
\eeq
where $D^h_q$ and $D^h_{\bar{q}}$ are the quark and antiquark fragmentation
functions and  $q(x),\bar q(x)$ are the corresponding parton
distribution functions. The ${\mathcal{O}}(\alpha_s)$ result for the
numerator in (\ref{bakwas}) can be expressed in terms of 
\begin{equation}
\begin{array}{rcl}
 F^h_i(x,z)    & = &  \sum_q e_q^2
\int_x^1 d\xi \int_z^1 d\zeta\\
       &   & \times\{K_{i,qq}(\xi,\zeta) q(x/\xi)D^h_q(z/\zeta)
        +K_{i,qg}(\xi,\zeta) q(x/\xi)D^h_g(z/\zeta)\\
       &   & + \;K_{i,qq}(\xi,\zeta) \bar q(x/\xi)D^h_{\bar q}(z/\zeta)
   +K_{i,qg}(\xi,\zeta) \bar q(x/\xi)D^h_g(z/\zeta)\\
       &   & + \;K_{i,gq}(\xi,\zeta) g(x/\xi)D^h_q(z/\zeta)
   +K_{i,gq}(\xi,\zeta) g(x/\xi)D^h_{\bar q}(z/\zeta)\}
\end{array}\label{ullu}
\end{equation}
where the $F^h_i$ denote the longitudinal or transverse pieces that
appear in the numerator on the RHS of (4). In the $K_{i,qq}$ for instance, the $i$ stands for the longitudinal or transverse
contribution according to what is required and the double suffix
denotes the contribution from an incoming as well as a fragmenting quark/anti-quark. Similarly the
other terms have corresponding suffixes that represent incoming gluons
(photon-gluon fusion) and fragmenting quarks/anti-quarks or incoming quarks/anti-quarks and
fragmenting gluons. The variable $\zeta$ denotes the longitudinal
momentum of the partons while $\xi =\frac{Q^2}{2p.q}$ with $p$ the
incoming parton momentum. In terms of the squared matrix elements
\beq\label{Kqq}
K_{i,qg}(\xi,\zeta) =\frac{\as}{2\pi}
C_F C_{i,q}(\xi,\xi-\xi\zeta)
\eeq 
and
\beq
K_{i,gq}(\xi,\zeta)=\frac{\as}{2\pi}
T_R C_{i,g}(\xi,\xi-\xi\zeta)
\;.
\eeq
while
\beq
K_{i,qq}(\xi,\zeta)=\frac{\as}{2\pi}
C_F C_{i,q}(\xi,1-\xi+\xi\zeta)
\eeq
where the $C_i$ represent the matrix elements squared \cite{myself}
for scattering off incoming quarks or gluons as denoted by the
suffixes($q$ or $g$). It is emphasised that we do not compute the contribution with
incoming gluons here. It is expected that for the kinematic range (small $x$
values) at HERA the above could be an important effect while perhaps
it is
not too significant for fixed target experiments. 
\section{Power corrections}
As mentioned earlier we calculate the power corrections based on the
dispersive approach. This means computing the first order correction
(\ref{ullu})
with a finite gluon mass $\epsilon$. The coefficient of the $\epsilon
\log{\epsilon}$ term then corresponds to the coefficient of the
$1/Q^2$ power behaved term. Computing (\ref{ullu}) with a massive
gluon matrix element and phase space and then taking double moments
with respect to $x$ and $z$ we can express the convolution in that
equation as a product involving double moments of  the $K$ functions and single
moments of parton distributions and fragmentation functions. Then
keeping only the leading non analytic terms in $\epsilon$ one finds  

\begin{equation}
\begin{array}{rcl}
\tilde\cK_{T,qq}(N,M,\eps)
&=&\left[2S_1(N)+ 2S_1(M+1)-3+\frac{1}{N}+\frac{1}{N+1}
+\frac{1}{M+1}+\frac{1}{M+2}\right]\ln\eps\nonumber\\ 
&&+\Biggl[-4S_1(N+1)-4S_1(M+1)+6+2N+2M+2NM \nonumber\\
&&-\frac{M+2}{N+1}-\frac{4}{N+2}-\frac{N+4}{M+1}
\Biggr]\eps\ln\eps\nonumber\\
\tilde\cK_{T,qg}(N,M,\eps)
&=& \left[-\frac{2}{M}+\frac{2}{M+1}
-\frac{1}{M+2}\right]\ln\eps + 
\left[1+\frac{4}{M}-\frac{N+4}{M+1}\right]\eps\ln\eps\nonumber\\
\tilde\cK_{L,qq}(N,M,\eps)
&=& \left[4-\frac{8}{N+2}\right]\eps\ln\eps\nonumber\\
\tilde\cK_{L,qg}(N,M,\eps) &=&0
\end{array}
\end{equation}

where

\beq
S_1(N) = \sum_{j=1}^{N-1} \frac 1 j\;.
\eeq
and
\beq
\tilde\cK_{i,qq}(N,M,\epsilon)=\int_0^1d\xi\int_0^1d\zeta \; \xi^N
\zeta^M \tilde\cK_{i,qq}(\xi,\zeta,\epsilon)
\eeq

The $\cK$ functions differ from the corresponding $K$ functions we had
in (\ref{ullu}) only by a factor which we choose to include in the
definition of our phenomenological parameter $\c{A}_2$.
Here we have included the virtual contribution calculated in \cite{BPY}.
The expression given above for the gluon fragmentation
contribution $\tilde\cK_{T,qg}$ is valid only for $M>2$.
There is an infrared divergence at $M=0$,
because we integrate the real gluon contribution
over one hemisphere only, which does not suffice to cancel
the divergent virtual contribution at $\zeta=0$.
For $M=1$ there is a contribution of $8\sqrt{\eps}$ instead
of an $\eps\ln\eps$ term, implying a $1/Q$-correction to this
moment of the gluon fragmentation function,
as is the case in $\ee$ annihilation \cite{self}.
For $M=2$ the 1 becomes --1 in the coefficient of $\eps\ln\eps$.
All of these changes represent extra contributions at the point
$\zeta=0$, which we can ignore because the fragmentation function
at any finite $z$ depends only on the behaviour at $\zeta>z>0$.

The $\ln\eps$ terms generate the
logarithmic scaling violations in the structure and fragmentation
functions while the $\eps\ln\eps$
terms give rise to $1/Q^2$ power
corrections, as mentioned earlier. We also need to consider power corrections in the
denominator of (\ref{bakwas}) and the results for this are taken from
an earlier publication \cite{DasWebDIS}. 
\section{Results and conclusions}
We assume in what follows that the quark fragmentation function is
independent of flavour and that $D_q=D_{\bar{q}}$ which should be reasonable if one confines the
discussion to the light quarks and if we sum over the fragmentation
into all charged particles.
One can combine the leading order results (\ref{Borncs}) with the
power corrections and express the results as
\begin{equation}
\label{Dhexpn}
\begin{array}{rcl}
F^h(z;x,Q^2) &=& D_q(z)+
\frac{\cA_2}{Q^2}\frac{1}{f(x)}
\int_x^1 d\xi \int_z^1 d\zeta\,f(x/\xi)\,
\{ [H_{T,qq}(\xi,\zeta)\nonumber\\
&&-H_{T,q}(\xi)\delta(1-\zeta)]D_q(z/\zeta)
 +H_{T,qg}(\xi,\zeta)D_g(z/\zeta) \}\;,
\end{array}
\end{equation}

where $f(x)$ is the charge-weighted parton distribution and $H_{T,q}(\xi)$ is the
higher-twist coefficient function for the transverse structure
function 
.\footnote{Note that the definition
here differs from that in Refs.~\cite{DasWebDIS} by a
factor of $-1/\xi$.}
\beq\label{HTq}
H_{T,q}(\xi) = \frac{4}{(1-\xi)_+}-2-4\xi
+4\delta(1-\xi)+\delta'(1-\xi)\;.
\eeq
The negative contribution from the structure function comes about
because of our normalization to the structure functions in equation
(\ref{bakwas}). Also the longitudinal contribution is identical between the
numerator and denominator of equation (\ref{bakwas}) which means it does not
appear in the final result.
We represent our final result as 
\beq\label{Dhres}
F^h(z;x,Q^2) = D_q(z)\left(1+\frac{\cA_2}{Q^2}\,H(z;x)\right)
\eeq
We plot the function $H(z,x)$ as a function of $z$ for different
values of $x$.

For the plot we use the ALEPH \cite{ALEPH} parametrizations of the light quark
and gluon fragmentation functions for charged hadrons at $Q=22$ GeV, and the
corresponding MRST (central gluon) \cite{MRS98} parton distributions.
Thus the predictions are at $Q^2 = 484$ GeV$^2$, but $H(z;x)$ depends
only weakly (logarithmically) on $Q^2$, and in any case our method is
not reliable at the level of logarithmic variations.
Results become insensitive to $x$ below the values shown
in Fig.1. Recall, however, that we have not computed the
singlet contribution, which may well be important at low $x$
because of the increase in the gluon distribution there. 

\begin{figure}
\begin{center}
\epsfig{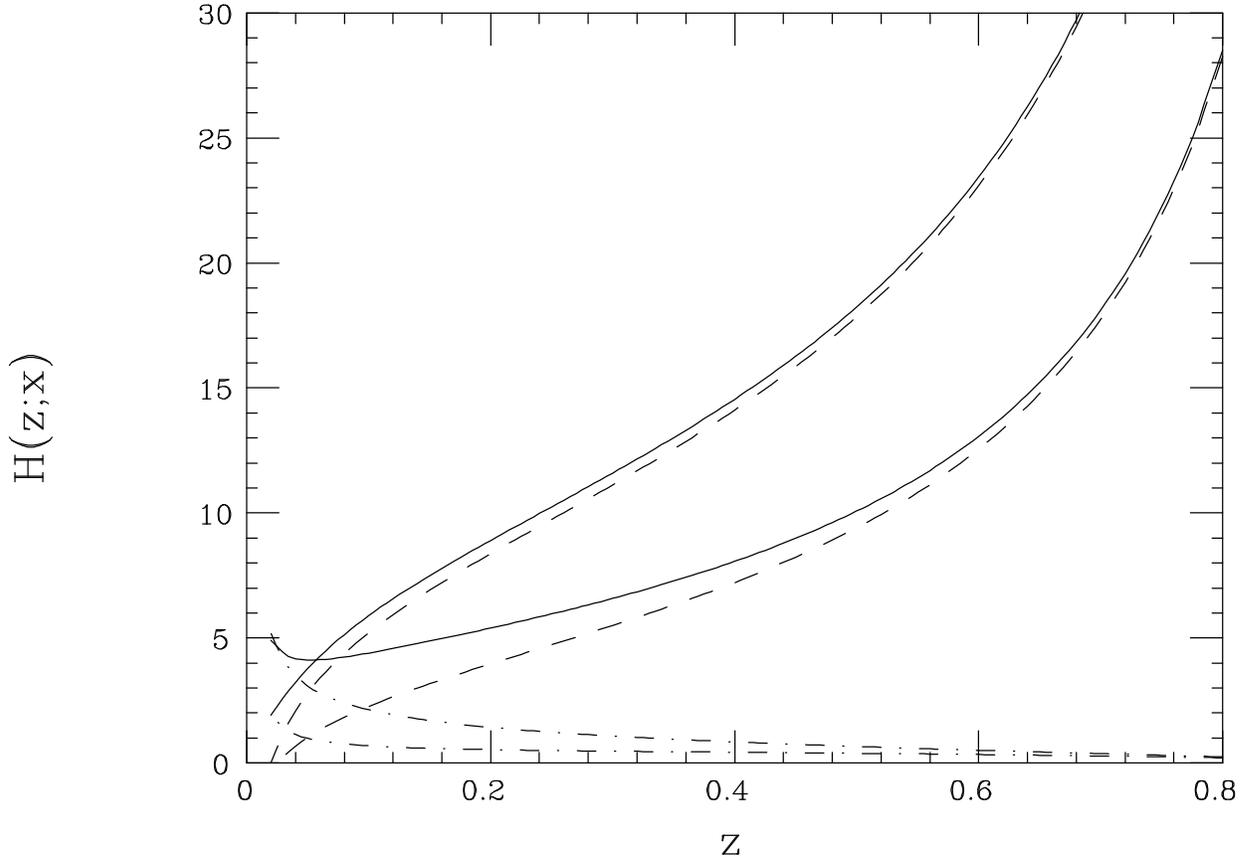}
\caption{Dashed, dot-dashed and solid curves are quark, gluon and
total fragmentation. The two sets of curves are for $x=0.1$(lower) and $x=0.3$(upper)
\label{fig:radish}}
\end{center}
\end{figure}
The predicted power corrections are qualitatively similar to those
for fragmentation functions in $\ee$ annihilation \cite{self},
though somewhat larger in magnitude. Part of the increase
comes from the negative higher-twist correction to the transverse
structure function in the denominator of Eq.(\ref{bakwas}). The contribution from gluon fragmentation,
although subject to further corrections \cite{myself}
is estimated to be relatively small for $z>0.2$.

The last point that needs to be made is that inclusion of electroweak
effects does not change the qualitative situation and in fact just
means redefining the function $f(x)$ to include electroweak couplings
rather than just the quark charges.
\section*{Acknowledgments}
The work presented here was carried out in collaboration with G.E.Smye
and B.R.Webber. I would like to thank Trinity College, the University
of Cambridge and
the organizing commitee of Moriond QCD for financial support.

\end{document}